\documentclass[english,a4paper,prl,twocolumn,amsmath,amssymb,superscriptaddress]{revtex4-1}

\renewcommand{\vec}[1]{\boldsymbol{\mathbf{#1}}}
\usepackage{pdfpages}
\makeatletter
\AtBeginDocument{\let\LS@rot\@undefined}
\makeatother
\usepackage{pgffor}

\begin{document}

\title{Morphometric approach to many-body correlations in hard spheres}
\date{\today}

\author{Joshua F. Robinson}
\email{joshua.robinson@bristol.ac.uk}
\affiliation{H.\ H.\ Wills Physics Laboratory, University of Bristol, BS8 1TL, UK}
\author{Francesco Turci}
\affiliation{H.\ H.\ Wills Physics Laboratory, University of Bristol, BS8 1TL, UK}
\author{Roland Roth}
\affiliation{Institut f\"ur Theoretische Physik, Universit\"at T\"ubingen, 72076 T\"ubingen, Germany}
\author{C. Patrick Royall}
\email{paddy.royall@bristol.ac.uk}
\affiliation{H.\ H.\ Wills Physics Laboratory, University of Bristol, BS8 1TL, UK}
\affiliation{School of Chemistry, Cantock’s Close, University of Bristol, BS8 1TS, UK}
\affiliation{Centre for Nanoscience and Quantum Information, Bristol BS8 1FD, UK}

\begin{abstract}
  We model the thermodynamics of local structures within the hard sphere liquid at arbitrary volume fractions through the \textit{morphometric} calculation of $n$-body correlations.
  We calculate absolute free energies of local geometric motifs in excellent quantitative agreement with molecular dynamics simulations across the liquid and supercooled liquid regimes.
  We find a bimodality in the density library of states where five-fold symmetric structures appear lower in free energy than four-fold symmetric structures, and from a single reaction path predict a relaxation barrier which scales linearly in the compressibility factor.
  The method provides a new route to assess changes in the free energy landscape at volume fractions dynamically inaccessible to conventional techniques.
\end{abstract}


\maketitle

\emph{Introduction.}---While mean-field theories provide insight into complex phenomena, physical accuracy is ensured only by a proper treatment of correlations.
For example, the simplest case of two-body correlations is at the foundation of predictive theories of the liquid state \cite{Hansen2013}, colloids and complex plasmas \cite{likos2001,Ivlev2012}.
In particular, the thermodynamics of simple liquids with solely pairwise interactions can be exactly expressed in terms of two-body correlations \cite{Hansen2013}. However, to resolve these integrated quantities \emph{spatially} into structural motifs, and \emph{temporally} into specific dynamical events, one needs to calculate many-body correlations.
While such a many-body approach may often be neglected in normal liquids, longstanding challenges such as the dramatic dynamical changes occurring in supercooled liquids approaching their glass transition \cite{Berthier2011,Royall2015} and phase transitions such as crystal nucleation \cite{Russo2012} call for a many-body description.

In the case of supercooled liquids, theories based on pair correlations such as the standard mode-coupling framework \cite{goetze} fail to account for activated events thus predicting a spurious ergodicity breaking transition \cite{Brambilla2009,Hallett2018}.
Activated dynamics are often rationalised through collective (i.e.\ many-body) effects within contrasting thermodynamic and purely dynamic scenarios \cite{Lubchenko2007,Tarjus2005,Biroli2006,Janssen2015,Szamel2013,Chandler2010}.
These include exact mean-field results in high dimensions \cite{Parisi2010,Charbonneau2017} whose relevance in finite-dimensional systems is hotly debated \cite{Wyart2017}.
A finite-dimensional theoretical description of many-body effects is therefore much needed.

However, many-body correlations are challenging to compute and typically combine both energetic and entropic contributions.
Physical insight can be gleaned by exploring the potential energy landscape of isolated clusters \cite{Wales2004,Arkus2009}, but such methods are only exhaustive for small system sizes.
This limitation has been partly addressed by embedding clusters in a mean-field approximation of the surrounding liquid \cite{Mossa2003}.
Nonetheless, this approach neglects by construction the intra-cluster entropic contributions that may dominate in the supercooled regime of interest.
Furthermore computer simulations, which naturally deliver full many-body correlations are limited in the range of dynamics they can access, hampering an approach to the glass transition, except for recent developments for certain models \cite{Berthier2016}.

Here we place theoretical predictions of many-body local structure on a fundamentally more rigorous footing using inhomogeneous liquid state theory \cite{Evans1979}.
We model the many-body interactions between a local subsystem and the remaining liquid, directly accessing the many-body \textit{free} energy of local arrangements of particles.
This allows us to predict the populations of specific local structures in the bulk system across the entire liquid phase and beyond the dynamically accessible supercooled regime.

\begin{figure}[b]
  \centering
  \includegraphics[width=75mm]{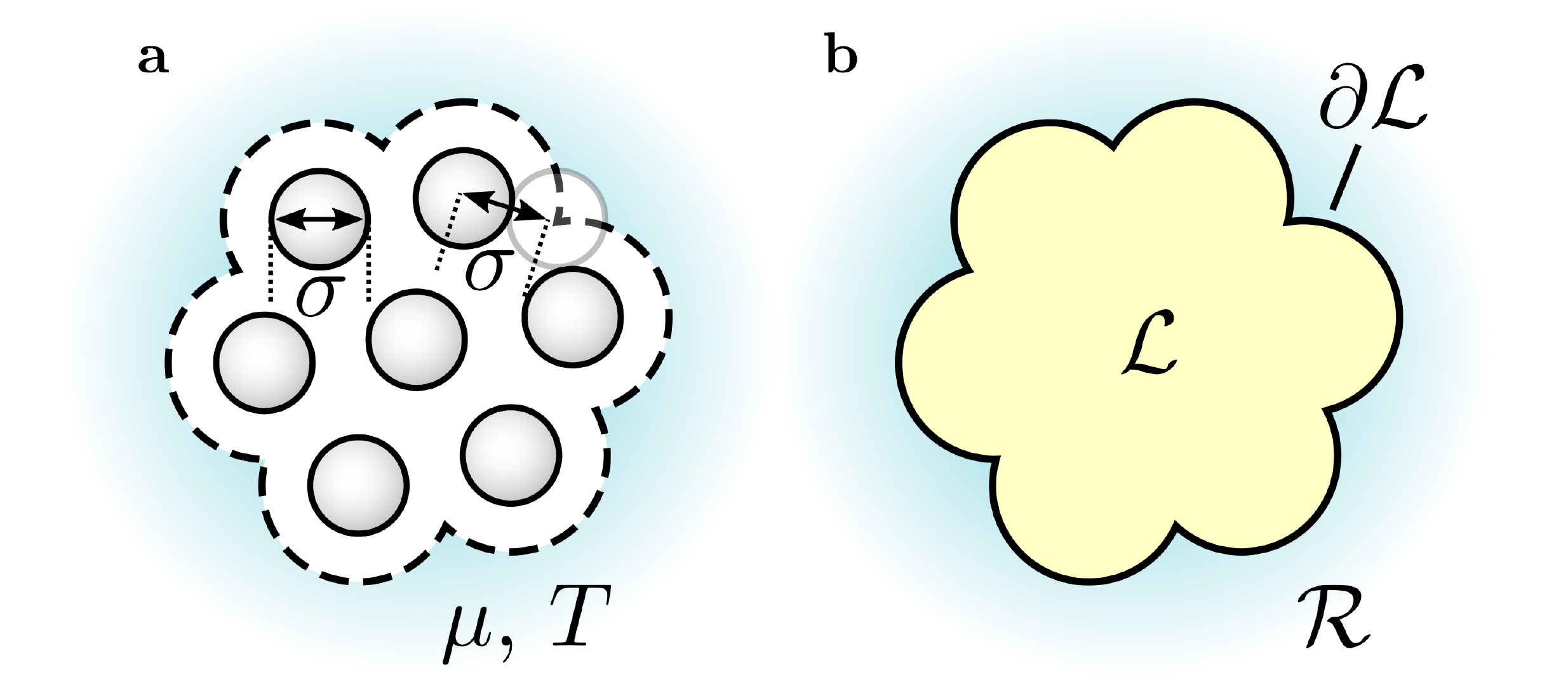}
  \caption{
    (color online)
    The system considered showing
    (a) the local particles surrounded by the remaining liquid acting as a thermal reservoir at fixed chemical potential and temperature, and
    (b) partition of space into the local $\mathcal{L}$ and remaining $\mathcal{R}$ components with dividing surface $\partial\mathcal{L}$.
    In this work $\mathcal{L}$ is chosen as the space inaccessible to the centre of a test particle (shown faded) representing the remaining liquid.
  }
  \label{fig:system}
\end{figure}

\emph{Many-body correlations and surface tension.}---We conceptually separate the liquid into $n$ spatially adjacent particles and the remaining degrees of freedom, acting as a solvent, which we treat within the grand-canonical ensemble, as sketched in Fig.\ \ref{fig:system}(a).
The joint probability density for simultaneously finding $n$ identical particles embedded in $\mathbb{R}^d$ at positions $\vec{r}^n := \{\vec{r}_1, \dots,\vec{r}_n\}$ is proportional to the \emph{$n$-particle distribution function} $g^{(n)}(\vec{r}^n)$ \cite{Hansen2013}.
For a homogeneous system, this can be formally expressed in terms of the \emph{potential of mean force}, the reversible work required to insert the particles at $\vec{r}^n$:
\begin{equation}\label{eq:potential-mean-force}
  \begin{split}
    \phi^{(n)}(\vec{r}^n) &\equiv - k_B T \ln g^{(n)}(\vec{r}^n) \\
    &= U(\vec{r}^n) + \Delta \Omega(\vec{r}^n) - n\mu^{ex}.
  \end{split}
\end{equation}
We denote by $U$ the total potential energy of the $n$ interacting particles and by $\Delta\Omega := \Omega - \Omega_{hom}$ the difference between the grand potential of the homogeneous liquid $\Omega_{hom}$ (related to the total volume and pressure by the relation $\Omega_{hom} = -pV$) and the grand potential of the system including the $n$-particle inhomogeneity. Finally, $k_B T$ and $\mu^{ex}$ are the thermal energy and the excess chemical potential (with respect to the ideal gas) of the homogeneous liquid respectively.

For systems with excluded volume interactions, we can divide the space into a local component $\mathcal{L} \subset \mathbb{R}^d$ of volume $V_\mathcal{L}$ inaccessible to solvent degrees of freedom, and the remaining space $\mathcal{R} = \mathbb{R}^d \setminus \mathcal{L}$ filled by solvent (Fig. \ref{fig:system}).
The dividing surface $\partial\mathcal{L}$ separates these two components with surface area $A_{\partial\mathcal{L}}$, creating a surface tension $\gamma$.
The solvent contribution to Eq.\ \eqref{eq:potential-mean-force} is then \begin{equation}\label{eq:surface-tension}
  \Delta \Omega[\mathcal{L}] =
  p V_\mathcal{L} + \gamma[{\partial\mathcal{L}}] A_{\partial\mathcal{L}}.
\end{equation}
Note that the surface tension is not unique as only the total grand potential must be independent of the choice of $\partial\mathcal{L}$ and can even change its sign for some choices of dividing surface \cite{Bryk2003}.
For simplicity we will consider one-component liquids with particles of diameter $\sigma$.
Letting $B_R(\vec{r}_i)$ denote a ball of radius $R$ at site $\vec{r}_i$, we choose the \emph{solvent accessible surface} \cite{Lee1971} as the dividing surface such that $\mathcal{L} = \cup_{i=1}^n B_\sigma(\vec{r}_i)$ (Fig.\ \ref{fig:system}).

\emph{Integral geometry approximation for surface tension.}---While approaches rooted in classical density functional theory \cite{Evans1992} would derive the surface tension $\gamma$ in terms of complex functionals for the grand potential $\Delta\Omega[\rho(\vec{r})]$ dependent on the solvent density profile \cite{Rosenfeld1989,Roth2010}, we directly expand $\gamma[\partial\mathcal{L}]$ in terms of the morphological properties of the dividing surface $\partial\mathcal{L}$.
With the use of theorems from integral geometry \cite{Hadwiger1957} we are able to dramatically reduce the computational cost of the calculation, and accurately predict correlations at very high densities representative of the metastable supercooled state.

Following \cite{Konig2004} we assume $\Delta\Omega$ is translation and rotation invariant, continuous (with respect to the Hausdorff metric) and additive.
Hadwiger's characterisation theorem \cite{Hadwiger1957} then ensures the surface tension adopts the so-called \emph{morphometric} form
\begin{equation}\label{eq:morphometric-surface-tension}
  \gamma[\partial\mathcal{L}] =
  \gamma_\infty +
  \frac{\kappa \, C_{\partial\mathcal{L}} + \overline{\kappa} \, X_{\partial\mathcal{L}}}
       {A_{\partial\mathcal{L}}},
\end{equation}
with integrated mean and Gaussian curvatures $C_{\partial\mathcal{L}}$ and $X_{\partial\mathcal{L}}$, and $\gamma_\infty,\kappa,\overline{\kappa}$ as thermodynamic coefficients to be determined.
$\gamma_\infty$ is the surface tension at a planar wall (i.e.\ the familiar macroscopic surface tension), whilst $\kappa$ and $\overline{\kappa}$ are ``bending energies'' accounting for curvature corrections occurring at small length scales.
These values are system (and state-point) dependent, but do not depend on the local geometry, making the linear form of Eq.\ \eqref{eq:morphometric-surface-tension} desirable for calculation.
While strictly an approximation, we motivate Eq.\ \eqref{eq:morphometric-surface-tension} from numerical studies where it has been found to be highly accurate below the freezing volume fraction in hard spheres \cite{Roth2006,Laird2012,Blokhuis2013,Urrutia2014,Hansen-Goos2014}, and from the early success of scaled particle theories \cite{Reiss1959,Reiss1960}.

\begin{figure*}
  \includegraphics[width=\linewidth]{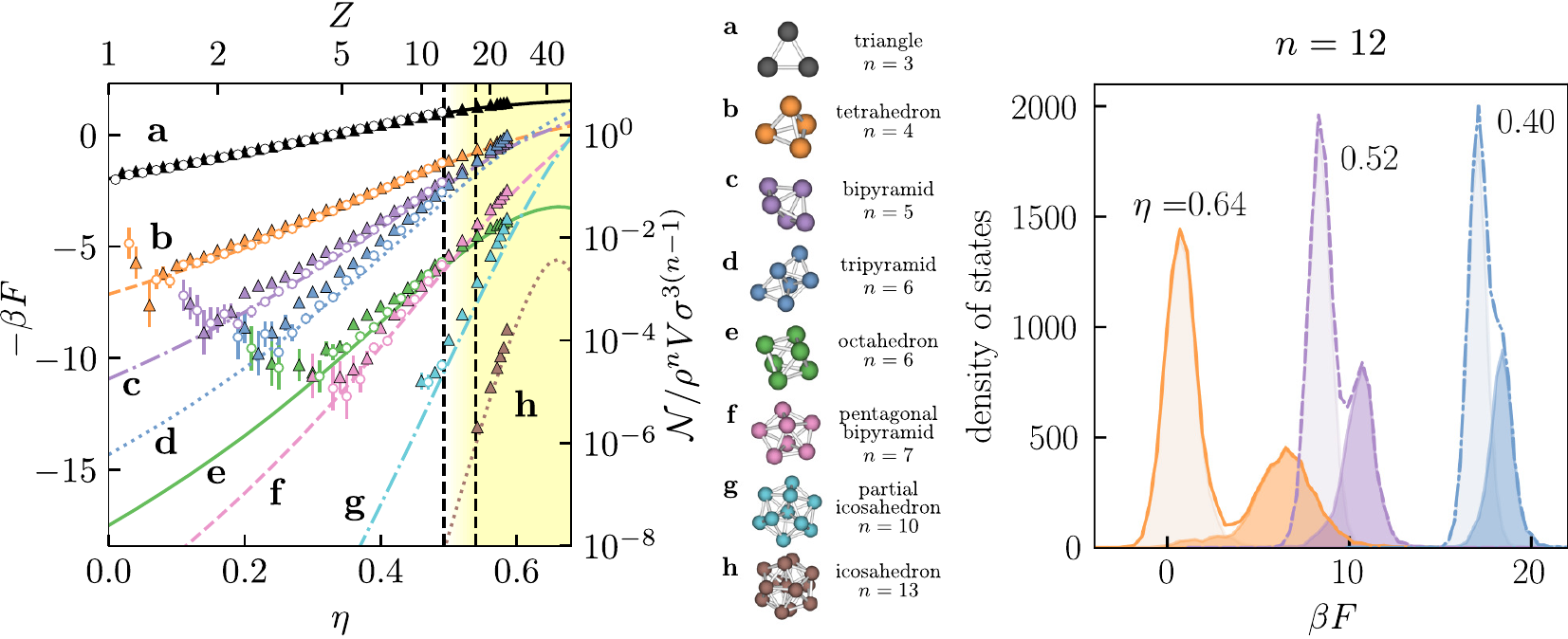}
  \caption{
    (color online)
    Static many-body structure in the hard sphere liquid.
    Left: populations of small local structures in the hard sphere liquid determined from molecular dynamics simulations of 1372 monodisperse (open circles) and 8\% polydisperse (solid triangles) hard spheres against the theoretical prediction of this work (lines).
    Variations against volume fraction $\eta$ and compressibility $Z = \beta p/\rho$ shown.
    The hard sphere freezing and melting volume fractions are indicated by vertical dashed lines.
    Right: theoretical free energy distribution for the $n=12$ local library of states at several volume fractions.
    The distribution is shifted to lower energies at higher volume fractions, and develops an increasingly bimodal structure.
    Populations are decomposed into those structures containing pentagonal bipyramids without octahedra (light fill) and the remaining structures (dark fill).
  }
  \label{fig:structure-populations}
\end{figure*}

\emph{Existing morphological theories.}---We focus on the hard sphere system because of its fundamental interest in the theory of liquids \cite{Widom1967,Hansen2013}.
This allows suitable coefficients of Eq.\ \eqref{eq:morphometric-surface-tension} to be derived analytically by exploiting the geometric nature of hard spheres.
We compute morphological quantities and their derivatives following \cite{Klenin2011}, which we have extended to calculate curvature measures (details in the Supplementary Material (SM)).
Note that hard spheres are athermal meaning density is the only control parameter and all free energies are really entropies; here we use ``supercooled'' to mean high density.

To proceed we need estimates of the thermodynamic coefficients $\gamma_\infty,\kappa,\overline{\kappa}$ accurate at high volume fractions and for typical $\partial\mathcal{L}$ morphologies.
The so-called White Bear II (WBII) theory provides coefficients \cite{Hansen-Goos2006} that are highly accurate in the limit of a planar $\partial\mathcal{L}$, however we find they predict inaccurate correlations for molecular geometries at densities above freezing.
In particular the contact value of $g^{(2)}$ with WBII coefficients spuriously decay to zero at the high densities of interest here (see SM).
For this reason we require a derivation of a new set of coefficients which we sketch below (full details in SM).
The derivation consists of a small modification to scaled particle theory  \cite{Reiss1959,Reiss1960} such that the virial theorem can be directly imposed,
\begin{equation}\label{eq:contact-g}
  g^{(2)}(\sigma) =
  \frac{3}{2\pi \sigma^3 \rho} \left( \frac{\beta p}{\rho} - 1 \right).
\end{equation}

\emph{Deriving new thermodynamic coefficients.}---We assume the Carnahan-Starling (CS) equation of state \cite{Carnahan1969} as this pressure is used in the WBII theory and is accurate deep within the supercooled regime \cite{Berthier2016} although it will fail at very large densities nearing random close packing.
We need three other equations to set the thermodynamic coefficients in Eq.\ \eqref{eq:morphometric-surface-tension} and obtain generic many-body correlations in the hard-sphere liquid.

First, by geometrical considerations \cite{Reiss1959}, we note that the cost of inserting a single hard point is exactly $\Delta \Omega[B_{\frac{\sigma}{2}}] = -k_B T \ln{(1- \eta)}$, where the (occupied) volume fraction is $\eta = \rho \pi \sigma^3 / 6$.
Second, the excess chemical potential is identically the cost of inserting an additional particle giving \cite{Widom1963} $\Delta \Omega[B_{\sigma}] = \mu^{ex}.$
The third equation comes by directly imposing the virial theorem \cite{Hansen2013} on the morphometric form of $g^{(2)}$ (Eqs.\ \eqref{eq:contact-g} and (4.15) in SM).
For two particles, the dividing surface $\partial\mathcal{L}$ resembles a ``dumbbell'' and the morphological quantities (and thus $g^{(2)}$ by Eqs.\ \eqref{eq:potential-mean-force} and \eqref{eq:morphometric-surface-tension}) have a simple form which can be calculated explicitly (see Ref.\ \cite{Oettel2009} and SM).
Solving the above expressions with the \emph{ansatz} \eqref{eq:morphometric-surface-tension} gives a new set of coefficients given explicitly in the SM.
The pair correlation produced by these coefficients is self-consistent with CS at contact by construction, moreover the new coefficients provide a theory that outperforms the older WBII approach across the whole range of distances typical of neighbouring particles (SM).
This enables us to accurately model complex many-particle local structures.

\emph{Free energy of local structures.}---Owing to the high accuracy of the correlations produced with the new morphometric coefficients, we can now calculate many-body correlations in the supercooled regime.
We denote the population of some chosen local structure as $\mathcal{N} = \rho^n V \sigma^{3(n-1)} e^{-\beta F}$ where $F$ is the free energy of the local structure.
From the definition of $g^{(n)}$ as a probability distribution we write the free energy as
\begin{equation}\label{eq:local-structure-free-energy}
  \beta F = -\ln{
    \frac{1}{V \sigma^{3(n-1)}}
    \left(
    \int_{\mathcal{D}}
    g^{(n)}(\vec{r}^n) \, d\vec{r}^n
    \right)
  },
\end{equation}
where the domain of integration $\mathcal{D}$ \emph{defines} the local structure, and $g^{(n)}$ is calculated from the morphometric potential of mean force using Eqs.\ \eqref{eq:potential-mean-force}, \eqref{eq:surface-tension} and \eqref{eq:morphometric-surface-tension} (computational details in SM).
We define a particular local structure by its bond topology, using a pairwise cutoff $\sigma_{cut}$ such that separations between particles are in the range $r_{ij} \in [\sigma, \sigma_{cut}]$ if they are ``bonded'' and $r_{ij} > \sigma$ otherwise.
All results presented use a cutoff of $\sigma_{cut}=1.2 \sigma$, but we have tested our findings are are not significantly affected by a choice of $\sigma_{cut}=1.4 \sigma$ indicating their robustness.

To demonstrate the effectiveness of this approach we have taken rigid structures for $3 \le n \le 13$ which are global minima of clusters in simple liquids \cite{Wales2004}.
We determined their free energies at arbitrary volume fraction by thermodynamic integration (details in SM) of Eq.\ \eqref{eq:local-structure-free-energy}.
In the left panel of Fig.\ \ref{fig:structure-populations} we find excellent agreement between the theoretical prediction and the observed concentration of local structure seen in molecular dynamics simulations of both mono- and moderately poly-disperse (8\%) hard spheres at all volume fractions accessed by the simulations i.e.\ $\eta \lesssim 0.585$ (details in SM).

Our approach is able to predict populations of local structures well beyond the regime dynamically accessible to simulation, finding nontrivial structural change deep in the glassy regime highlighted by a rescaling with respect to the trivial $\rho^n$ density contribution.
The free energy of considered structures changes approximately linearly across the entire liquid regime, with deviations from linear becoming more apparent in the supercooled regime.

All structures apart from the four-fold symmetric octahedron in Fig.\ \ref{fig:structure-populations} are subunits of the icosahedron, and increase in concentration more rapidly than the octahedron until high density.
For $n=6$ we consider the free energies of two structures: the tripyramid and octahedron.
We find that the tripyramid occurs $\sim20$ times more often than the octahedron, their free energy difference being dominated by the different point group symmetries following \cite{Malins2009,Meng2010}.
We can also estimate vibrational contributions, which allow us to match not only the relative but also the absolute values of free energies obtained from simulation.
In particular, we are able to capture the gradual reduction of the population of octahedral motifs in favour of the tripyiramids at high volume fractions.
This is related to the previously observed emergence of five-fold symmetric motifs (such as the full and partial icosahedron) \cite{Royall2015,Tarjus2005,Hallett2018,Dunleavy2015} which is here directly predicted from liquid state theory.

Having tested that the theory is accurate for selected geometries, we now take the exhaustive list of 11980 rigid structures for $n=12$ determined in \cite{Holmes-Cerfon2016} to obtain a local density of states for a given sized inhomogeneity.
These rigid structures correspond to unique contact topologies, but in thermal systems (i.e. with finite gaps between particles) we expect many of them to be indistinguishable as found in Ref.\ \cite{Trombach2018}.
Nevertheless, due to their exhaustiveness these represent a complete local density of states in the liquid, of fundamental interest to random first--order transition theory \cite{Lubchenko2007}.
We calculated the free energy of all (first-order) rigid (nonsingular) structures using Eq.\ \eqref{eq:local-structure-free-energy} (right panel Fig.\ \ref{fig:structure-populations}), finding a bimodal distribution with two main peaks separated by a free energy difference that increases with increasing volume fraction.
We find the that lower energy distribution consists of structures rich in five-fold (icosahedral) symmetry in the absence of four-fold (octahedral) symmetry.

\begin{figure}
  \includegraphics{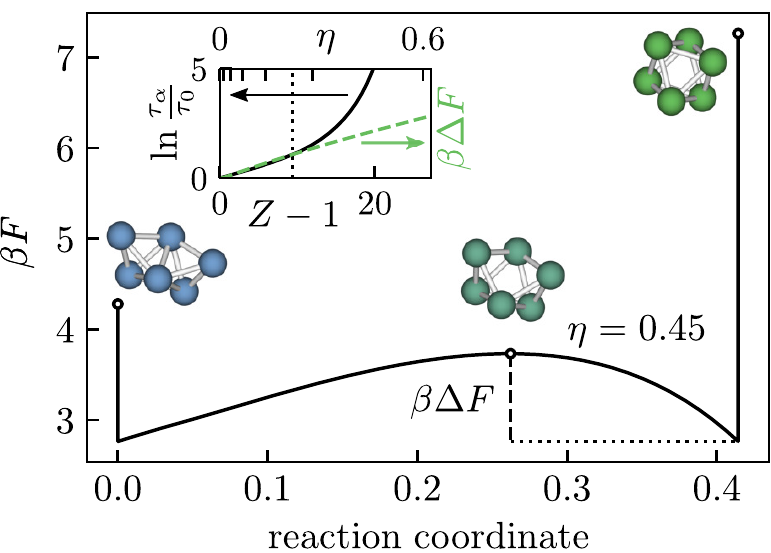}
  \caption{
    (color online)
    Reaction for transition between tripyramid and octahedron $n = 6$ structures.
    Stationary points are indicated by markers: there is a discontinuity in free energy at the end points due to the additional integration over the reaction coordinate, and symmetry in the case of the octahedron.
    Inset: variation of activation barrier with volume fraction $\eta$ and compressibility $Z = \beta p/\rho$ from this theoretical reaction path (dashed line) and measured $\alpha$--relaxation times in bulk molecular dynamics simulations (solid line), where $\eta = 0.45$ is indicated with a vertical dotted line.
  }
  \label{fig:reaction}
\end{figure}

\emph{Dynamics: free energy along a reaction path.}---We have thus far focused on static thermodynamic properties: yet a connection with dynamics can be made by calculating the free energy along reaction paths between (geometrically similar) structures.
This calculation along unstable directions in the free energy landscape requires an analytic approach (described in the SM), and generates paths such as the one in Fig.\ \ref{fig:reaction}.
Here we consider transitions between the tripyramid and the octahedron with $n=6$ as this is the simplest nontrivial transition between distinct hard sphere packings (SM).
Comparing this dynamical barrier to the structural relaxation for ($\alpha$--) relaxation timescale $\tau_\alpha$ extracted from simulations relative to a microscopic time $\tau_0$ (inset of Fig.\ \ref{fig:reaction}), we find this single reaction path barrier agrees with the low density scaling of $\tau_\alpha$ (linear in the compressibility factor $Z$ \cite{Berthier2009}).
However, activated dynamics are not expected in this regime so this agreement may be coincidental.
It is possible to extend our methodology for larger rearrangements, which may be sufficient to access ($\alpha$--) relaxation \emph{at very deep supercooling} for equilibrium systems.
However, the rapid growth in the number of possible states presents a considerable numerical challenge requiring new methods and approximations, so we leave this exciting avenue for future study.

\emph{Conclusions.}---We have presented a formalism for describing many-body correlations in liquids and developed it into an accurate and computationally efficient parameter-free theory for hard spheres using integral geometry relying solely on the choice of the equation of state.
The key approximations involved treating the grand potential as continuous and additive (related to extensivity), and imposing the correct contact value of $g^{(2)}(r)$.

We applied the framework to a selection of local structural correlations, therefore predicting nontrivial changes in the energy landscape with supercooling putting previous empirical observations on more solid ground.
In particular, our analysis provides evidence for the existence of two populations of structures with distinct symmetries and free energies which causes the local density of states to become increasingly bimodal at high densities.
We note that we have treated densities corresponding to a degree of supercooling only accessible using novel swap Monte-Carlo techniques \cite{Berthier2016}; however, these simulations introduce large polydispersity, changing the local structure \cite{Coslovich2018} and thus limiting direct comparison with our calculations for the monodisperse liquid.

Our framework can be easily adapted to more complex liquids such as systems with soft repulsive interactions and polydisperse mixtures \cite{Kodama2011}.
Integral geometry underlies the core equation \eqref{eq:morphometric-surface-tension}, so this approach can extend to hard particles of more complex shapes where the interaction potential is still geometric in nature.
It is applicable to a more general class of liquids where the soft part of the potential may be treated as a perturbation around a hard core \cite{Hansen2013} such that a geometric decomposition still applies.
This suggests a new route for predicting static properties of equilibrium liquids, with direct applications to self-assembly, nucleation and protein structure.

\begin{acknowledgements}
  \emph{Acknowledgements.}---We are indebted to Bob Evans for countless conversations which shaped the liquid state foundations of this work, and for carefully reading the manuscript.
  We are grateful to Chiara Cammarota, Daniele Coslovich, Giuseppe Foffi and Gilles Tarjus for stimulating discussions concerning the relevance to supercooled liquids.
  JFR, FT and CPR  acknowledge the European Research Council under the FP7 / ERC Grant agreement n$^{\circ}$617266 ``NANOPRS''. CPR would like to acknowledge the Royal Society for financial support.
\end{acknowledgements}

%

\newpage
\onecolumngrid


\section*{Supplementary material}

# Supplementary material for "Morphometric approach to many-body correlations in hard spheres"

Joshua F. Robinson,[1, *] Francesco Turci,[1] Roland Roth,[2] and C. Patrick Royall[1, 3, 4, †]

[1]*H. H. Wills Physics Laboratory, University of Bristol, UK*
[2]*Institut für Theoretische Physik, Universität Tübingen, 72076 Tübingen, Germany*
[3]*School of Chemistry, Cantock's Close, University of Bristol, BS8 1TS, UK*
[4]*Centre for Nanoscience and Quantum Information, Bristol BS8 1FD, UK*

## CONTENTS

## I. GENERALISED POTENTIAL OF MEAN FORCE FROM GRAND CANONICAL ENSEMBLE

We write the $n$-particle distribution function $g^{(n)}$ as

$$\text{Probability } (any \ n \text{ particles in volume } d\mathbf{r}^n) \equiv \rho^n g^{(n)}(\mathbf{r}^n) \, d\mathbf{r}^n. \tag{1.1}$$

In the main text, we expressed $g^{(n)}$ in terms of a generalised potential of mean force $\phi^{(n)}$: the reversible work required to insert $n$ particles into the liquid. We decomposed $\phi^{(n)}$ into a local (potential energy) and solvent (free energy) component. Although this quantity is quite intuitive and could be determined heuristically, here we give a short proof that this decomposition is formally exact and arises quite naturally from the definition of the distribution function.

---

\* joshua.robinson@bristol.ac.uk
† paddy.royall@bristol.ac.uk

In the grand-canonical ensemble the $n$-particle density $\rho^{(n)}(\mathbf{r}^n) \equiv \rho^n g^{(n)}$ is determined by integrating over the remaining degrees of freedom [1]

$$\rho^{(n)}(\mathbf{r}^n) = \frac{1}{\Xi} \sum_{N=n}^{\infty} \frac{z^N}{(N-n)!} \int e^{-\beta U_N} \, d\mathbf{r}^{(N-n)}, \tag{1.2}$$

where the activity is written in terms of the thermal de Broglie wavelength $\Lambda$ as $z = \exp \beta \mu / \Lambda^d$. Changing the summation limits $N \to N + n$ we obtain

$$\rho^{(n)}(\mathbf{r}^n) = \frac{z^n}{\Xi} \sum_{N=0}^{\infty} \frac{z^N}{N!} \int e^{-\beta U_{N+n}} \, d\mathbf{r}^N \tag{1.3}$$

$$= z^n e^{-\beta U_n} \left\langle e^{-\beta U_{n \leftrightarrow N}} \right\rangle$$

where in the latter step we decomposed the total potential $U_{N+n}$ into purely local and solvent terms, i.e. $U_{N+n} = U_n + U_N + U_{n \leftrightarrow N}$, where $U_\alpha$ for $\alpha \in \{n, N\}$ indicates the internal interactions between particles in component $\alpha$. The "interspecies" interactions are contained within $U_{n \leftrightarrow N}$ which acts as an external field for the solvent. The angled brackets indicate ensemble averaging over all arrangements of the solvent, i.e.

$$\langle \cdots \rangle = \frac{1}{\Xi} \sum_{N=0}^{\infty} \frac{z^N}{N!} \int (\cdots) \, e^{-\beta U_N} \, d\mathbf{r}^N, \tag{1.4}$$

with partition function of the unperturbed system $\Xi \equiv e^{-\beta \Omega_{hom}}$, where $\Omega_{hom} = -pV$ is the usual homogeneous grand potential. Thus, (1.3) becomes

$$\rho^{(n)}(\mathbf{r}^n) = z^n e^{-\beta (U_n + \Omega - \Omega_{hom})}. \tag{1.5}$$

where $\Omega$ is the grand potential of the solvent in the presence of the $n$-particle inhomogeneity. Splitting the chemical potential into its ideal and excess parts so that $\beta \mu = \ln \Lambda^d \rho + \beta \mu^{ex}$ gives

$$\rho^{(n)}(\mathbf{r}^n) = \rho^n e^{-\beta (U_n + \Omega - \Omega_{hom} - n\mu^{ex})}. \tag{1.6}$$

The $n$-particle distribution functions are then determined from [1]

$$g^{(n)}(\mathbf{r}^n) \equiv \frac{\rho^{(n)}(\mathbf{r}^n)}{\rho^n} = e^{-\beta (U_n + \Delta\Omega - n\mu^{ex})} \tag{1.7}$$

where $\Delta\Omega = \Omega - \Omega_{hom}$ giving the generalised potential of mean force stated in the main text.

The above is essentially the generalisation of the *potential distribution theorem* [2, 3] to many-particles. See Ref. [4] and references therein for a detailed review of this approach.

## II. INTEGRAL CURVATURE MEASURES AND THEIR GRADIENTS

The two curvature measures, $C_{\partial \mathcal{L}}$ and $X_{\partial \mathcal{L}}$ (integrated mean and Gaussian curvatures respectively), are required for the morphological thermodynamics used extensively in the main text. In this section we give details on their geometric construction to aid morphological calculations. The relevant formulas for $C_{\partial \mathcal{L}}$ and $X_{\partial \mathcal{L}}$ in hard sphere systems have previously been given in references [5] and [6]. Here, we restate these formulas and extend them by computing their derivatives with respect to atomic coordinates. This technical description is only likely to be of interest to those wishing to do morphological calculations of their own.

To briefly motivate these derivative calculations, we remind the reader that derivatives were used in the main text in the calculation of the free energy of local structures along the octahedron-tripyramid reaction path. Gradients were required for this calculation, providing an analytic method (with perturbation theory) where the numerical method (thermodynamic integration) fails due the instability of intermediate points along the reaction path. Full details of this method are given in Sec. V D. It is worth stating that the usefulness of gradient calculations extends beyond this one application.

The gradient gives the mean depletion forces between (nearby) particles within the bulk liquid, which is generally a quantity of interest in liquid state theories. These solvation forces are useful for speeding up numerical minimisation procedures, and for describing the solvation forces for molecules and proteins in aqueous solution. For the latter case one requires a solvent accessible surface $\partial \mathcal{L}$ which is composed of balls of varying radii; the formulas we present allow for this generalisation.

## A. Decomposing the solvent accessible surface into intersections

For correlations in homogeneous liquids composed of identical balls, the curvatures must be computed across the solvent accessible surface $\partial \mathcal{L}$ where the enclosed volume is

$$\mathcal{L} = \cup_{i=1}^n B_\sigma(\mathbf{r}_i).$$

In order to keep the formulas as general as possible, we will consider a small generalisation of this surface where the spheres are of arbitrary radii, i.e.

$$\mathcal{L} = \cup_{i=1}^n B_{\sigma_i}(\mathbf{r}_i), \tag{2.1}$$

where $\sigma_i$ is the diameter of particle $i$. One obtains the surface used in the main text by setting $\sigma_i = \sigma$ for all $i$. Let $K^d \subseteq \mathbb{R}^d$ denote the space of *polyconvex* subsets of $d$-dimensional Euclidean space. By *polyconvex* we mean subsets composed of a *finite* union of convex subsets; this means the surfaces are well-behaved so standard geometric descriptions and intuitions apply. By construction $\mathcal{L} \in \mathcal{K}^d$ and $\partial \mathcal{L} \in \mathcal{K}^{d-1}$.

If particle $i$ is on the surface of $\mathcal{L}$, i.e.

$$S_i \cap \partial \mathcal{L} \notin \emptyset$$

where $S_i \equiv \partial B_{\sigma_i}(\mathbf{r}_i)$ is the spherical surface of particle $i$, then integrals over $\partial \mathcal{L}$ must carefully consider pieces $S_i$ and intersections $\bigcap_i S_i$ separately. Intersections, e.g. $S_i \cap S_j$ for $i \neq j$, contribute zero area, but may have nonvanishing curvature; this is usually understood by considering the parallel surface $\partial(\mathcal{L} + B_\epsilon)$ in the limit as $\epsilon \to 0$. Hard core interactions ensure pathological cases where spheres share a centre are excluded, so intersections of two spheres must result in a (one) lower dimensional manifold $S_i \cap S_j \in \mathcal{K}^{d-2}$. It is straightforward to extend this argument to $n \leq d$ intersections $\bigcap_{i=1}^n S_i \in \mathcal{K}^{d-n}$. For $n = d$ intersections the solution is a zero-dimensional manifold, i.e. a point.

Intersections between $n > d$ spherical surfaces are possible in principle, but in practice they occur with vanishing probability once a system is thermalised. To see this, consider an overlap of $n > d$ hard spheres. By the above argument one can decompose the surface of the resulting structure into $k$-dimensional submanifolds where $0 \leq k \leq d$. Higher order surface intersections $n > d$ requires multiple $n = d$ intersections to occur at the same point, which is overconstrained and occurs with measure zero. Thus, the probability of finding points where $n > d$ spheres intersect occurs with measure zero. This argument only applies because we considered the *boundary* of the intersections; it is more common in the physics literature to consider the Mayer f-function (related to the Euler characteristic) of the intersection *volumes*, where the measure is nonzero leading to slow convergence of the virial series [1]. Despite having zero measure, there are cases where one would construct a geometry containing a higher order intersection so we will return to this topic in more detail in section II D.

In summary, for $d = 3$ the surface $\partial \mathcal{L}$ contains the following submanifolds:

- $S_i \cap \partial \mathcal{L} \notin \emptyset$: a spherical cap from particle $i$.

- $S_i \cap S_j \cap \partial \mathcal{L} \notin \emptyset$ for $i \neq j$: a line, specifically a circular arc.

- Point $S_i \cap S_j \cap S_k \cap \partial \mathcal{L} \notin \emptyset$ for $i \neq j \neq k$: points where balls $i$, $j$, and $k$ intersect. 3 intersecting spheres generally have 2 points of intersection, though usually only 1 of these coincides with the surface $\partial \mathcal{L}$ (the other is usually buried inside the volume $\mathcal{L}$).

- Intersections of more than 3 surface spheres: occurs with vanishing probability in *thermal* systems (see above).

In subsequent sections we will detail contributions to $C_{\partial \mathcal{L}}$ and $X_{\partial \mathcal{L}}$ (and their derivatives) from each of these intersections in $d = 3$. Note that for morphological thermodynamics one also requires the volume and surface area contributions, $V_\mathcal{L}$ and $A_{\partial \mathcal{L}}$, for which we do *not* provide computational details as there is already a wealth of literature on this subject (notably Refs. [7, 8]). In particular, we found the algorithm in Ref. [9] to be fast and robust. We extended their implementation to also compute integrated curvature measures, with the formulas given in subsequent sections.

## B. Integrated mean curvature

### 1. Notation

In the main text we expressed the coordinates $\mathbf{r}^n \in \mathbb{R}^{3n}$ as the direct sum $\mathbf{r}^n = \mathbf{r}_1 \oplus \cdots \oplus \mathbf{r}_n$. As each line contribution in (2.2) depends on the positions of 2 central spheres, whose intersection forms a circle, together with

2 additional spheres whose intersections with the first two spheres creates terminating vertices for the circular arc. Thus the line contribution depends on the positions of up to 4 particles, so the domain and image of the gradient is (potentially) 12-dimensional. Fewer particles can be involved, and thus the dimension of the space is reduced, if:

- $\theta_l^{(1)}$ and $\theta_l^{(2)}$ are formed by intersections with the same third particle, i.e. if both the solutions (points) to $S_1 \cap S_2 \cap S_i$ $i \notin \{1, 2\}$ and the line joining them is on the surface $\partial S$. In this case the space is 9-dimensional.

- $\phi_l = 2\pi$: the line forms a closed circle, uninterrupted by other particle intersections. In this case the space is 6-dimensional.

These cases are pathological and generally lines involve 4 particles so the space is 12-dimensional.

Given the large dimensionality we adopt the following notation to carefully distinguish each term in the space. We recast the coordinates as the product space $\mathbf{r}^n \in \mathbb{R}^{n \times 3}$ so we can use the following notation for basis vectors:

$$\mathbf{e}_i^\alpha = \mathbf{e}_i \otimes \mathbf{e}_\alpha,$$

where $i \in \{1, \ldots, n\}$ and $\alpha \in \{1, 2, 3\}$. We use Roman indices to denote the particle number and Greek indices for the Cartesian component.

We use suffix notation where summation is assumed over repeated indices. For example, we write the coordinates in summation convention as

$$\mathbf{r}^n = r_i^\alpha \mathbf{e}_i^\alpha := \sum_{i,\alpha} r_i^\alpha \mathbf{e}_i^\alpha.$$

One obtains the gradient from components of differentiation by summing over the basis vectors, as in

$$\boldsymbol{\nabla} f = \frac{\partial f}{\partial r_i^\alpha} \mathbf{e}_i^\alpha.$$

The latter expression can be used to determine gradients from the explicit forms of differentials given in subsequent sections.

## 2. Problem

The integrated mean curvature for the generalised solvent accessible surface $\partial \mathcal{L}$ where $\mathcal{L}$ is composed of spheres of differing radii as in (2.1) is given as [6]

$$C_{\partial \mathcal{L}} = \sum_{s \in S} \frac{A_s}{\sigma_s} - \sum_{l \in L} \frac{\phi_l R_l}{2} (\theta_l^{(1)} + \theta_l^{(2)}), \tag{2.2}$$

where $A_s$ is the area of spherical cap $s$, $\phi_l$ is the angular length of the line $l$, $\phi_l R_l$ its arc length, and $\theta_l^{(1)}$ and $\theta_l^{(2)}$ are the angles between the spheres and the plane of intersection defined in Refs. [10, 11], see Fig. 1(a). Algorithms already exist for computing the differential of the total area [8, 9], so we do not need to consider the first term in (2.2). This leaves only the line contributions for consideration. The curvature contribution from a single line differentiates to give

$$\frac{\partial C_{\partial \mathcal{L}}}{\partial r_i^\alpha} = \sum_{s \in S} \frac{1}{\sigma_s} \frac{\partial A_s}{\partial r_i^\alpha} - \frac{1}{2} \sum_{l \in L} \left( \frac{\partial \phi_l}{\partial r_i^\alpha} R_l (\theta_l^{(1)} + \theta_l^{(2)}) + \phi_l \frac{\partial R_l}{\partial r_i^\alpha} (\theta_l^{(1)} + \theta_l^{(2)}) + (\phi_l R_l) \left( \frac{\partial \theta_l^{(1)}}{\partial r_i^\alpha} + \frac{\partial \theta_l^{(2)}}{\partial r_i^\alpha} \right) \right). \tag{2.3}$$

In subsequent sections we will give explicit formulas for each differential in this expression.

To get the curvature for the surface used in the main text one sets $\sigma_i = \sigma \ \forall \ i \in \{1 \ldots n\}$. In this limiting case the curvature (2.2) simplifies to

$$C_{\partial \mathcal{L}} = \frac{A_{\partial \mathcal{L}}}{\sigma} - \sum_{l \in L} \frac{\phi_l R_l}{2} (\theta_l^{(1)} + \theta_l^{(2)}),$$

and many of the explicit expressions for $\phi_l$, $R_l$, $\theta_l^{(1)}$ and $\theta_l^{(2)}$ also simplify as will be seen in subsequent sections.

Note that in the computational algorithm which we use to compute $\partial \mathcal{L}$ (c.f. Ref. [9]) the symbol $\theta$ denotes the in-plane angle $\frac{\pi}{2} - \theta_l^{(\alpha)} \ \forall \ \alpha \in \{1, 2\}$ is used instead in the construction of the surface, so formulas stated below must be adjusted in the implementation if this angle is used.

### 3. Particle separations

The separation between two particle centers is

$$\boldsymbol{\Delta}_{ij} = \mathbf{r}_i - \mathbf{r}_j \tag{2.4}$$

which differentiates to

$$\frac{\partial |\boldsymbol{\Delta}_{ij}|}{\partial r_k^\alpha} = \frac{r_i^\alpha - r_j^\alpha}{|\boldsymbol{\Delta}_{ij}|}(\delta_{ki} - \delta_{kj}) \tag{2.5}$$

where $\delta_{ij}$ is the Kronecker delta.

The basis vector for separations differentiates as

$$\frac{\partial}{\partial r_k^\alpha}\left(\frac{\boldsymbol{\Delta}_{ij}}{|\boldsymbol{\Delta}_{ij}|}\right) = \frac{(\delta_{ki} - \delta_{kj})\mathbf{e}_k^\alpha}{|\boldsymbol{\Delta}_{ij}|} - \frac{\boldsymbol{\Delta}_{ij}}{|\boldsymbol{\Delta}_{ij}|^2}\frac{\partial |\boldsymbol{\Delta}_{ij}|}{\partial r_k^\alpha}. \tag{2.6}$$

### 4. Quantities in the plane orthogonal to intersection

We will label the particles whose intersection creates the circle of arc as 1 and 2 for convenience. We consider the distances from these particles to the center of the circle as $b_l^{(1)}$ and $b_l^{(2)}$, see Fig. 1(a). Clearly $|\boldsymbol{\Delta}_{12}| = b_l^{(1)} + b_l^{(2)}$. These distances form equilateral triangles with circle radius $R_l$ of hypotenuse $\sigma_i$ for $i \in \{1, 2\}$. This geometry is sketched in Fig. 1a. By Pythagoras' theorem we find the unknown distances as

$$R_l = \sigma_1 \cos\theta_l^{(1)} = \sigma_2 \cos\theta_l^{(2)} = \frac{1}{2}\sqrt{2(\sigma_1^2 + \sigma_2^2) - |\boldsymbol{\Delta}_{12}|^2 - \left(\frac{\sigma_1^2 - \sigma_2^2}{|\boldsymbol{\Delta}_{12}|}\right)^2}, \tag{2.7}$$

$$b_l^{(1)} = \sigma_1 \sin\theta_l^{(1)} = \frac{|\boldsymbol{\Delta}_{12}| + \frac{\sigma_1^2 - \sigma_2^2}{|\boldsymbol{\Delta}_{12}|}}{2}, \tag{2.8}$$

$$b_l^{(2)} = \sigma_2 \sin\theta_l^{(2)} = \frac{|\boldsymbol{\Delta}_{12}| - \frac{\sigma_1^2 - \sigma_2^2}{|\boldsymbol{\Delta}_{12}|}}{2}. \tag{2.9}$$

The gradients of the angles between the planes are

$$\frac{\partial \theta_l^{(1)}}{\partial r_i^\alpha} = \frac{1 - \frac{\sigma_1^2 - \sigma_2^2}{|\boldsymbol{\Delta}_{12}|^2}}{2R_l}\frac{\partial |\boldsymbol{\Delta}_{12}|}{\partial r_i^\alpha}, \tag{2.10a}$$

$$\frac{\partial \theta_l^{(2)}}{\partial r_i^\alpha} = \frac{1 + \frac{\sigma_1^2 - \sigma_2^2}{|\boldsymbol{\Delta}_{12}|^2}}{2R_l}\frac{\partial |\boldsymbol{\Delta}_{12}|}{\partial r_i^\alpha}, \tag{2.10b}$$

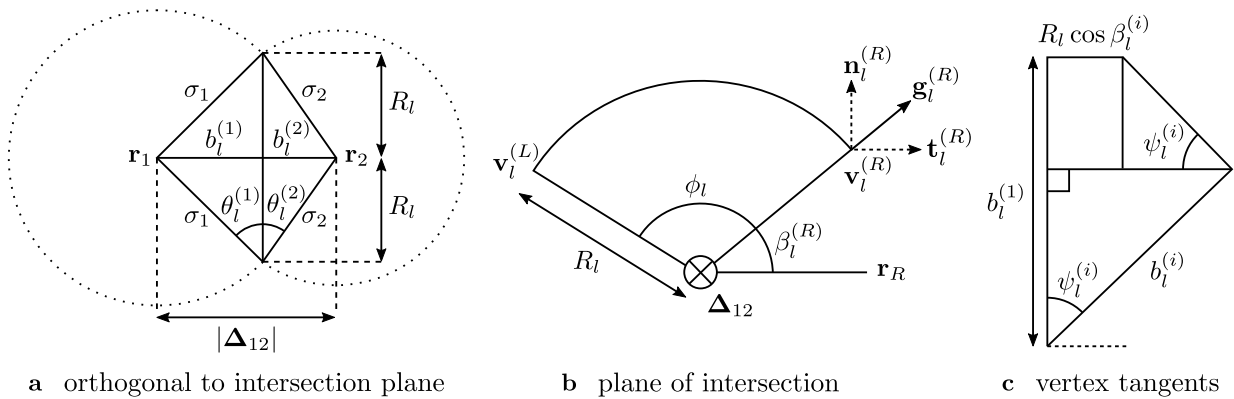

**a** orthogonal to intersection plane      **b** plane of intersection      **c** vertex tangents

FIG. 1. Geometrical quantities involved in the calculation of line curvatures.

and the gradients of the distances are

$$\frac{\partial R_l}{\partial r_i^\alpha} = \frac{|\boldsymbol{\Delta}_{12}|}{4 R_l} \left( \left( \frac{\sigma_1^2 - \sigma_2^2}{|\boldsymbol{\Delta}_{12}|} \right)^2 - 1 \right) \frac{\partial |\boldsymbol{\Delta}_{12}|}{\partial r_i^\alpha}, \tag{2.11a}$$

$$\frac{\partial b_l^{(1)}}{\partial r_i^\alpha} = \frac{1 - \frac{\sigma_1^2 - \sigma_2^2}{|\boldsymbol{\Delta}_{12}|^2}}{2} \frac{\partial |\boldsymbol{\Delta}_{12}|}{\partial r_i^\alpha}, \tag{2.11b}$$

$$\frac{\partial b_l^{(2)}}{\partial r_i^\alpha} = \frac{1 + \frac{\sigma_1^2 - \sigma_2^2}{|\boldsymbol{\Delta}_{12}|^2}}{2} \frac{\partial |\boldsymbol{\Delta}_{12}|}{\partial r_i^\alpha}. \tag{2.11c}$$

As we must calculate the angles $\theta_l^{(1)}$ and $\theta_l^{(2)}$ and their derivatives for the curvature calculation, i.e. Eqs. (2.2) and (2.3), a convenient form for these latter derivatives is

$$\frac{\partial R_l}{\partial r_i^\alpha} = -b_l^{(1)} \frac{\partial \theta_l^{(1)}}{\partial r_i^\alpha} = -b_l^{(2)} \frac{\partial \theta_l^{(2)}}{\partial r_i^\alpha}, \tag{2.12a}$$

$$\frac{\partial b_l^{(1)}}{\partial r_i^\alpha} = R_l \frac{\partial \theta_l^{(1)}}{\partial r_i^\alpha}, \tag{2.12b}$$

$$\frac{\partial b_l^{(2)}}{\partial r_i^\alpha} = R_l \frac{\partial \theta_l^{(2)}}{\partial r_i^\alpha}. \tag{2.12c}$$

The expressions in this section are the only quantities which explicitly depend on the sizes of the spheres $\sigma_i$, so we see how it is straightforward to consider the more general surface composed of arbitrarily sized spheres. The above formulas are simplified if spheres are of equal sizes i.e. $\sigma_i = \sigma$ (as in the main text) leading to vanishing of $\sigma_1^2 - \sigma_2^2$ terms.

## 5. Angular length

To complete the derivatives in (2.3) we need an explicit expression for the gradient of the angular separation $\phi_l$. In general, the line consists of an arc along a circle of radius $R_l$, which terminates at the vertices.

We consider the 2d plane containing the intersection $S_1 \cap S_2$. The center of the intersection circle is

$$\mathbf{c}_l = \mathbf{r}_1 + b_l^{(1)} \frac{\boldsymbol{\Delta}_{12}}{|\boldsymbol{\Delta}_{12}|} = \mathbf{r}_2 - b_l^{(2)} \frac{\boldsymbol{\Delta}_{12}}{|\boldsymbol{\Delta}_{12}|}. \tag{2.13}$$

The plane of intersection is seen by looking along $\boldsymbol{\Delta}_{12}$, making the arc $\phi_l R_l$ perfectly circular, shown in Fig. 1b. By convention we choose the arc to be a clockwise migration from a 'left' vertex to a 'right' vertex, which we label $v_l^{(L)}$ and $v_l^{(R)}$ accordingly. These vertices form at 3-particle intersections, so we will use the suffixes $L$ and $R$ to indicate the 3rd particle index (distinct from 1 and 2) where appropriate. We write the vertex coordinates as

$$\mathbf{v}_l^{(i)} = \mathbf{c}_l + R_l\, \mathbf{g}_l^{(i)} \qquad i \in \{L, R\}, \tag{2.14}$$

where the unit vectors are

$$\mathbf{g}_l^{(i)} = \frac{\mathbf{v}_l^{(i)} - \mathbf{c}_l}{|\mathbf{v}_l^{(i)} - \mathbf{c}_l|} \qquad i \in \{L, R\}. \tag{2.15}$$

The unit vectors for the 'right' component are sketched in Fig. 1b. The central angle is thus defined as the angle (going clockwise) between these unit vectors, i.e.

$$\cos \phi_l = \mathbf{g}_l^{(L)} \cdot \mathbf{g}_l^{(R)}, \tag{2.16}$$

which after differentiation gives

$$\frac{\partial \phi_l}{\partial r_i^\alpha} = -\frac{\frac{\partial \mathbf{g}_l^{(L)}}{\partial r_i^\alpha} \cdot \mathbf{g}_l^{(R)} + \mathbf{g}_l^{(L)} \cdot \frac{\partial \mathbf{g}_l^{(R)}}{\partial r_i^\alpha}}{\sin \phi_l}. \tag{2.17}$$

So we need explicit expressions for vectors $\mathbf{g}_l^{(L)}$ and $\mathbf{g}_l^{(R)}$ and their derivatives to proceed.

We decompose the vertex unit vectors into the following Cartesian basis:

$$\mathbf{g}_l^{(i)} = \cos \beta_l^{(i)} \, \mathbf{t}_l^{(i)} + \sin \beta_l^{(i)} \, \mathbf{n}_l^{(i)} \qquad i \in \{L, R\}, \tag{2.18}$$

where $\mathbf{t}_l^{(i)}$ is the vector tangent to the plane spanned by particles $\{1, 2, i\}$ and $\mathbf{n}_l^{(i)}$ is normal to this plane. If $\psi_l^{(i)}$ is the angle $\angle 1i2$, that is

$$\cos \psi_l^{(i)} = \frac{\boldsymbol{\Delta}_{12} \cdot \boldsymbol{\Delta}_{1i}}{|\boldsymbol{\Delta}_{12}||\boldsymbol{\Delta}_{1i}|} \qquad i \in \{L, R\} \tag{2.19}$$

then these basis vectors take the form

$$\mathbf{n}_l^{(L)} = \frac{1}{\sin \psi_l^{(L)}} \frac{\boldsymbol{\Delta}_{12} \times \boldsymbol{\Delta}_{1L}}{|\boldsymbol{\Delta}_{12}||\boldsymbol{\Delta}_{1L}|}, \tag{2.20a}$$

$$\mathbf{n}_l^{(R)} = \frac{1}{\sin \psi_l^{(R)}} \frac{\boldsymbol{\Delta}_{1R} \times \boldsymbol{\Delta}_{12}}{|\boldsymbol{\Delta}_{1R}||\boldsymbol{\Delta}_{12}|}, \tag{2.20b}$$

$$\mathbf{t}_l^{(L)} = \frac{\mathbf{n}_l^{(L)} \times \boldsymbol{\Delta}_{12}}{|\boldsymbol{\Delta}_{12}|}, \tag{2.20c}$$

$$\mathbf{t}_l^{(R)} = \frac{\boldsymbol{\Delta}_{12} \times \mathbf{n}_l^{(R)}}{|\boldsymbol{\Delta}_{12}|}. \tag{2.20d}$$

Finally, from Fig. 1c we have

$$b_l^{(i)} \sin \beta_l^{(i)} - R_l \cos \beta_l^{(i)} = \frac{b_l^{(1)} - b_l^{(i)} \cos \psi_l^{(i)}}{\tan \psi_l^{(i)}} \qquad i \in \{L, R\}, \tag{2.21}$$

giving

$$\mathbf{g}_l^{(i)} \cdot \mathbf{t}_l^{(i)} = \cos \beta_l^{(i)} = \frac{1}{R_l} \left( b_l^{(i)} \sin \psi_l^{(i)} + \frac{b_l^{(i)} \cos \psi_l^{(i)} - b_l^{(1)}}{\tan \psi_l^{(i)}} \right) \qquad i \in \{L, R\}. \tag{2.22}$$

With explicit expressions for all of the vectors involved in the arc, the only remaining step is to differentiate. First, we differentiate the vertex unit vectors Eq. (2.18) in the basis defined by Eq. (2.20) to obtain

$$\frac{\partial \mathbf{g}_l^{(j)}}{\partial r_i^\alpha} = \frac{\mathbf{g}_l^{(j)} \times \boldsymbol{\Delta}_{12}}{|\boldsymbol{\Delta}_{12}|} \frac{\partial \beta_l^{(j)}}{\partial r_i^\alpha} + \cos \beta_l^{(j)} \frac{\partial \mathbf{t}_l^{(j)}}{\partial r_i^\alpha} + \sin \beta_l^{(j)} \frac{\partial \mathbf{n}_l^{(j)}}{\partial r_i^\alpha} \qquad j \in \{L, R\}. \tag{2.23}$$

Second, we take the derivatives of the basis vectors themselves Eq. (2.20) giving

$$\frac{\partial \mathbf{n}_l^{(L)}}{\partial r_i^\alpha} = \frac{\mathbf{e}_{c,j} \times \mathbf{e}_l + \mathbf{e}_c \times \mathbf{e}_{l,j}}{\sin \psi_l} - \frac{\cos \psi_l}{\sin \psi_l} \psi_{l,j} \, \mathbf{n}_l, \tag{2.24a}$$

$$\frac{\partial \mathbf{n}_l^{(R)}}{\partial r_i^\alpha} = \frac{\mathbf{e}_{r,j} \times \mathbf{e}_c + \mathbf{e}_r \times \mathbf{e}_{c,j}}{\sin \psi_r} - \frac{\cos \psi_r}{\sin \psi_r} \psi_{r,j} \, \mathbf{n}_r, \tag{2.24b}$$

Finally, we differentiate the angles from Eqs. (2.19) and (2.22) to get

$$\frac{\partial \beta_l^{(j)}}{\partial r_i^\alpha} = \frac{\frac{\partial R_l}{\partial r_i^\alpha} \cos \beta_l^{(j)} - \frac{\partial}{\partial r_i^\alpha} \left( R_l \cos \beta_l^{(j)} \right)}{R_l \sin \beta_l^{(j)}}, \tag{2.25a}$$

$$\frac{\partial}{\partial r_i^\alpha} \left( R_l \cos \beta_l^{(j)} \right) = \frac{1}{\sin \psi_l^{(j)}} \frac{\partial b_l^{(j)}}{\partial r_i^\alpha} - \frac{1}{\tan \psi_l^{(j)}} \frac{\partial b_l^{(1)}}{\partial r_i^\alpha} - \frac{b_l^{(j)} \cos \psi_l^{(j)} - b_l^{(1)}}{\sin^2 \psi_l^{(j)}} \frac{\partial \psi_l^{(j)}}{\partial r_i^\alpha}, \tag{2.25b}$$

$$\frac{\partial \psi_l^{(j)}}{\partial r_i^\alpha} = -\frac{1}{\sin \psi_l^{(j)}} \frac{\partial (\cos \psi_l^{(j)})}{\partial r_i^\alpha} = -\frac{1}{\sin \psi_l^{(j)}} \left( \frac{\partial}{\partial r_i^\alpha} \left( \frac{\boldsymbol{\Delta}_{12}}{|\boldsymbol{\Delta}_{12}|} \right) \cdot \frac{\boldsymbol{\Delta}_{1j}}{|\boldsymbol{\Delta}_{1j}|} + \frac{\boldsymbol{\Delta}_{12}}{|\boldsymbol{\Delta}_{12}|} \cdot \frac{\partial}{\partial r_i^\alpha} \left( \frac{\boldsymbol{\Delta}_{1j}}{|\boldsymbol{\Delta}_{1j}|} \right) \right), \tag{2.25c}$$

for $j \in \{L, R\}$ in each expression.

### C. Integrated Gaussian curvature

$\partial\mathcal{L}$ forms a closed two-dimensional surface, so by the Gauss-Bonnet theorem [12] the integrated Gaussian curvature must be

$$X_{\partial\mathcal{L}} = 2\pi\chi \tag{2.26}$$

where $\chi$ is the Euler characteristic of $\partial\mathcal{L}$, a topological constant of the surface. Thus, the derivative of $X$ is zero everywhere except at pathological points where a topological change in the surface occurs. In practice this happens when a cavity larger than a particle size forms inside the structure, or when a particle dissociates; being interested in *compact* local structures, we can exclude both of these scenarios from consideration. Thus, for all local structures $\chi = 2$, and the gradient of $X$ is zero everywhere.

Under the above assumptions we do not need to compute $X_{\partial\mathcal{L}}$, but nevertheless it is convenient to do so in order to check the correctness of the algorithm:

$$X_{\partial\mathcal{L}} = -\sum_{l\in L} \phi_l(\sin\theta_l^{(1)} + \sin\theta_l^{(2)}) + \sum_{(i,j,k)\in V} \Omega_{ijk} \tag{2.27}$$

where $\Omega_{ijk}$ is the solid angle spanned by the 3-vectors at vertex $S_i \cap S_j \cap S_k$ [5]. The condition that this sum must produce the same result as (2.26) provides a useful check of the numerics.

### D. Intersections of more bodies: possible caveats with quenched geometries

A common task with computer simulations is to quench a geometry to find the minimal (or *inherent*) energy structure. Quenched geometries are inherently athermal so the argument above that intersections of $n > d$ spheres should occur with vanishing probability does not apply; it is common to find intersections of 4 or more particles at the surface in $d = 3$ *after a quench*. To investigate quenched geometries one must properly treat such intersections, as we will demonstrate below with an illustrative example.

An example of these intersections in $d = 3$ occurs where 4 particles are arranged in a perfect square (as in e.g. any outer face of the body-centered cubic unit cell). This special geometry corresponds to a bifurcation point in configuration space, where two pairs of surface vertices are simultaneously created and annihilated. The gradient is not continuous at these points *with respect to the atomic coordinates*, as the continuity of the morphological free energy is only guaranteed with respect to the set $\mathcal{L}$ according to the Hausdorff metric. In atomic coordinates derivatives can contain poles and step discontinuities.

As the example above illustrates, the gradient is not well defined at pathological points where many spheres intersect. It is thus impossible to quench these geometries using standard algorithms (which assume smooth functions) and perturbation theories (e.g. the harmonic approximation for the free energy) will fail. A proper treatment of these cases would be required for an investigation of quenched geometries. In this work we have avoided these considerations primarily by restricting ourselves to geometries thermalised using a Monte-Carlo algorithm (Figures 2 and 3 in the main text). Construction of the reaction path of $n = 6$ particles necessitates an analytic method which uses a quenched geometry, however the geometries for $n = 6$ are not pathological so no special consideration is needed.

## III. EXPLICIT MORPHOLOGY FOR TWO PARTICLES

In the previous section we gave computational details for calculating morphological quantities of the solvent accessible surfaces. Here we give the explicit form for the special case where there are two particles. These formulas can provide some intuition for the general case, and will be directly used in section IV to derive new thermodynamic coefficients.

For two particles $g^{(2)}(\mathbf{r}_1, \mathbf{r}_2)$ reduces to $g^{(2)}(|\mathbf{r}_1 - \mathbf{r}_2|)$ as the system is completely isotropic. All morphological quantities are then functions of $r = |\mathbf{r}_1 - \mathbf{r}_2|$. As $r \to 2\sigma$ the solvent accessible surface $\partial\mathcal{L}$ self-intersects, and two separate (perfectly spherical) surfaces form for $r > 2\sigma$. The Euler characteristic of $\partial\mathcal{L}$ is thus

$$\chi(r) = \begin{cases} 2 & r < 2\sigma \\ 4 & r > 2\sigma. \end{cases} \tag{3.1}$$

Written explicitly, the resulting distribution function is

$$g^{(2)}(r) = \begin{cases} 0 & r < \sigma \\ \exp\Big(-\beta(pV(r) + \gamma_\infty A(r) + \kappa C(r) + \overline{\kappa}X(r) - 2\mu^{ex})\Big) & \sigma \le r \le 2\sigma \\ 1 & r > 2\sigma. \end{cases} \tag{3.2}$$

with morphological quantities

$$V(r) = \frac{8\pi}{3}\sigma^3 - (r^2 + 4\sigma r)\frac{\pi(2\sigma - r)^2}{12r}\,\Theta(2\sigma - r), \tag{3.3a}$$

$$A(r) = 8\pi\sigma^2 - 2\pi\sigma\,(2\sigma - r)\,\Theta(2\sigma - r), \tag{3.3b}$$

$$C(r) = 8\pi\sigma - 2\pi\left[\sqrt{\sigma^2 - \left(\frac{r}{2}\right)^2}\arcsin\left(\frac{r}{2\sigma}\right) + (2\sigma - r)\right]\Theta(2\sigma - r), \tag{3.3c}$$

$$X(r) = 2\pi\chi(r) \tag{3.3d}$$

where $\Theta(\cdots)$ is the Heaviside step function. The mean curvature stated is a special case of the more general result worked out in [13]. The first term of the expressions for $V, A, C$ contains the morphological measures for two *independent* particles (e.g. twice the volume of a single particle), whilst the remaining terms are corrections due to their mutual intersections.

## IV. DERIVATION OF MORPHOMETRIC COEFFICIENTS

### A. Hard sphere equation of state: Carnahan-Starling (CS)

The excess free energy is determined from the equation of state by

$$\frac{\beta F^{ex}}{N} = \int_0^\eta \left(\frac{\beta p}{\rho} - 1\right)\frac{d\eta'}{\eta'}, \tag{4.1}$$

giving the excess chemical potential from the thermodynamic relation

$$\beta\mu^{ex}[p] = \beta\left(\frac{\partial F^{ex}}{\partial N}\right)_{V,T} = \left(\frac{\beta p}{\rho} - 1\right) + \int_0^\eta \left(\frac{\beta p}{\rho} - 1\right)\frac{d\eta'}{\eta'}. \tag{4.2}$$

The Carnahan-Starling equation of state approximates the pressure for hard spheres as [14]

$$\frac{\beta p_{cs}}{\rho} = \frac{1 + \eta + \eta^2 - \eta^3}{(1 - \eta)^3}, \tag{4.3}$$

which gives the excess chemical potential using (4.2) as

$$\beta\mu_{cs}^{ex} = \frac{8\eta - 9\eta^2 + 3\eta^3}{(1 - \eta)^3}. \tag{4.4}$$

### B. Scaled particle route

We describe here in more detail the derivation of the first set of morphometric coefficients, equivalent to those given in [15] through fundamental measure theory (FMT). The derivation sketched below avoids use of FMT, instead favouring a geometric formulation equivalent to the scaled particle approach of Reiss *et al.* [16, 17]. The standard scaled particle approach considers an expansion of the grand potential surrounding a spherical solute in powers of radii; here, we modify the *ansatz* to use morphological measures instead, so that the resulting theory is more naturally extended to geometries of arbitrary shapes. Additionally, we impose the Carnahan-Starling equation of state as an *input* whereas the Percus-Yevick equation of state is an *output* of standard scaled particle approaches.

Following the protocol of scaled particle theories, we consider the insertion of a hard ball of radius $R - \frac{\sigma}{2}$ into the liquid at the origin. This choice of radius ensures that contact with the *center* of solvent particles occurs at distance

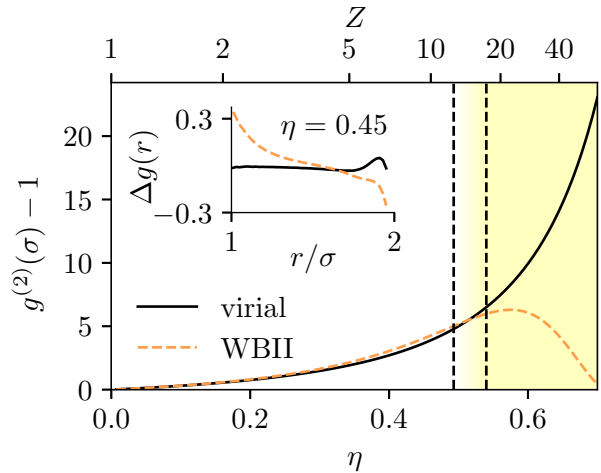

FIG. 2.    (color online) Contact values of the radial distribution function against volume fraction $\eta$ and reduced pressure $Z = \beta p / \rho$ for the hard sphere liquid using Eq. (4) from the main text with Eq. (3.2) for the explicit form of $g^{(2)}$, assuming the Carnahan-Starling equation of state. Contact values are determined with two sets of morphometric coefficients: WBII (section IV B and Ref. [15]) which performs poorly in the supercooled regime, and coefficients derived in this work (section IV C) using the virial theorem which is exact by construction. The hard sphere freezing and melting volume fractions are indicated by vertical dashed lines. Inset: the errors $\Delta g = g^{(2)} - g^{(2)}_{MD}$ where $g^{(2)}_{MD}$ is determined from molecular dynamics simulations (c.f. section VI), showing the new theory also improves the accuracy away from contact.

$R$ from the point of insertion i.e. $\rho(r) = 0$ for $r < R$. Writing the change in the grand potential due to the insertion of the ball in its morphometric form (from (2) and (3) of the main text), we have the *ansatz*

$$\Delta\Omega(R) = \frac{4\pi R^3}{3} p + 4\pi R^2 \, \gamma_\infty + 4\pi R \, \kappa + 4\pi \, \overline{\kappa}.  \tag{4.5}$$

If an equation of state for the pressure is taken as input, only three equations are needed to set the remaining coefficients of surface tension $\gamma_\infty, \kappa$ and $\overline{\kappa}$.

Restating the expressions for the insertion of a hard point and a new particle from the main text as

$$\Delta\Omega(R = 0) = -k_B T \ln(1 - \eta),  \tag{4.6}$$

$$\Delta\Omega\left(R = \frac{\sigma}{2}\right) = \mu^{ex},  \tag{4.7}$$

we need one more equation to set the thermodynamic coefficients for the theory. Following Ref. [18] we take the normal derivative of $\Omega$ with respect to $R$, and noting that $\Delta\Omega(R) = \Omega(R) - \Omega_{hom}$ gives

$$\left(\frac{\partial \Delta\Omega}{\partial R}\right)_{\mu, V, T} = \left(\frac{\partial \Omega}{\partial R}\right)_{\mu, V, T} = \int \frac{\delta\Omega[\rho_0(\mathbf{r})]}{\delta\rho} \left(\frac{\partial \rho_0(\mathbf{r})}{\partial R}\right)_{\mu, V, T} d\mathbf{r} + \int \rho_0(\mathbf{r}) \left(\frac{\partial \phi_{ext}(\mathbf{r})}{\partial R}\right)_{\mu, V, T} d\mathbf{r},  \tag{4.8}$$

where $\rho_0$ is the equilibrium density profile and $\phi_{ext}$ is the external potential (i.e. the potential of the ball). In equilibrium $\Omega$ is minimised so

$$\left.\frac{\delta\Omega[\rho(\mathbf{r}); \phi_{ext}]}{\delta\rho}\right|_{\rho(\mathbf{r}) = \rho_0(\mathbf{r})} = 0,  \tag{4.9}$$

so the first integral in (4.8) vanishes. As the ball is hard, the external potential and its derivative are zero everywhere except at the surface $R$ where both $\rho_0$ and $\phi_{ext}$ are discontinuous. We consider its Boltzmann weight, i.e.

$$e^{-\beta\phi_{ext}(\mathbf{r})} = \Theta(|\mathbf{r}| - R).  \tag{4.10}$$

Taking the derivative of both sides gives

$$\beta\left(\frac{\partial \phi_{ext}(\mathbf{r})}{\partial R}\right)_{\mu, V, T} = \delta(|\mathbf{r}| - R) e^{\beta\phi_{ext}(\mathbf{r})}  \tag{4.11}$$

Inserting this expression into (4.8) and using the fact that $\rho(\mathbf{r})e^{\beta\phi_{ext}(\mathbf{r})}$ is continuous (c.f. Ref. [1]) gives the contact theorem

$$\beta\left(\frac{\partial\Omega}{\partial R}\right)_{\mu,V,T} = 4\pi R^2\rho(R). \tag{4.12}$$

When $R = \sigma$ the inserted ball is equivalent to the hard sphere particles themselves, so $\Omega = \mu^{ex}$ and the contact density is $\rho(\sigma) = \rho\,g^{(2)}(\sigma)$ giving

$$\beta\left(\frac{\partial\Delta\Omega}{\partial R}\right)_{\mu,V,T}\bigg|_{R=\sigma} = \beta\left(\frac{\partial\Omega}{\partial R}\right)_{\mu,V,T}\bigg|_{R=\sigma} = 4\pi\sigma^2\rho\,g^{(2)}(\sigma), \tag{4.13}$$

or written in morphometric form using (4.5) we have

$$4\pi\sigma^2\,p + 8\pi\sigma\,\gamma_\infty + 4\pi\,\kappa = \frac{4\pi\sigma^2\rho}{\beta}\,g^{(2)}(\sigma). \tag{4.14}$$

Applying the virial theorem (equation (4) in the main text) to the right hand side gives the final expression:

$$4\pi\sigma^2\,p + 8\pi\sigma\,\gamma_\infty + 4\pi\,\kappa = \frac{6}{\beta\sigma}\left(\frac{\beta p}{\rho} - 1\right). \tag{4.15}$$

Together (4.6), (4.7) and (4.15) form a complete system of equations which we solve to obtain the coefficients

$$\frac{\beta\gamma_\infty^{WBII}}{R\rho} = \left(\frac{\pi}{6\eta^2} - \frac{5\pi}{18\eta}\right)p - \frac{\mu^{ex}[p]}{3\eta} - \frac{\ln(1-\eta)}{3\eta} - \frac{1}{\eta} \tag{4.16a}$$

$$\frac{\beta\kappa^{WBII}}{R^2\rho} = \left(\frac{4\pi}{9\eta} - \frac{\pi}{2\eta^2}\right)p + \frac{4\mu^{ex}[p]}{3\eta} + \frac{4\ln(1-\eta)}{3\eta} + \frac{3}{\eta} \tag{4.16b}$$

$$\frac{\beta\bar{\kappa}^{WBII}}{R^3\rho} = \left(\frac{\pi}{3\eta^2} - \frac{2\pi}{9\eta}\right)p - \frac{\mu^{ex}[p]}{\eta} - \frac{4\ln(1-\eta)}{3\eta} - \frac{2}{\eta}. \tag{4.16c}$$

Inserting the Carnahan-Starling parameters (4.3) and (4.4) gives the coefficients explicitly as

$$\frac{\beta p^{WBII}}{\rho} = \frac{1+\eta+\eta^2-\eta^3}{(1-\eta)^3} \tag{4.17a}$$

$$\frac{\beta\gamma_\infty^{WBII}}{R\rho} = -\frac{1+2\eta+8\eta^2-5\eta^3}{3(1-\eta)^3} - \frac{\ln(1-\eta)}{3\eta} \tag{4.17b}$$

$$\frac{\beta\kappa^{WBII}}{R^2\rho} = \frac{4-10\eta+20\eta^2-8\eta^3}{3(1-\eta)^3} + \frac{4\ln(1-\eta)}{3\eta} \tag{4.17c}$$

$$\frac{\beta\bar{\kappa}^{WBII}}{R^3\rho} = -\frac{4-11\eta+13\eta^2-4\eta^3}{3(1-\eta)^3} - \frac{4\ln(1-\eta)}{3\eta}, \tag{4.17d}$$

which are *identical* to the coefficients derived from the WBII free energy functional in Ref. [15]. Remarkably, we have obtained these coefficients through a route completely different from their original derivation. In Ref. [15] the coefficients were determined within FMT by taking the limit of a binary mixture where one component is infinitely dilute. Here we completely avoided FMT, in favour of geometrical arguments similar to standard scaled particle approaches. This suggests that the above scaled particle argument is somehow built into the structure of the WBII functional; we note that this is a nonobvious fact which cannot be determined from the form of the functional alone, nor is it obvious how it emerges from its original derivation.

Finally, note that (as described in the main text) the resulting $g^{(2)}$ performs poorly in the supercooled regime as compared with the "exact" result from virial theorem i.e. Eq. (4) in the main text. In Fig. 2 we plot the contact value with this set of coefficients, finding that it is reasonably accurate until around the freezing density where contact correlations spuriously decay. The next section will detail how to modify the derivation to produce coefficients which describe more accurate correlation functions at high densities.

### C. Virial route

In this section we give the detailed steps used in the derivation of the new set of morphometric coefficients, following the direction laid out in the main text. In section III above the morphological form of $g^{(2)}$ was computed explicitly; this derivation imposes the contact density using $\rho(\sigma) = \rho\, g^{(2)}(\sigma)$ with this explicit form.

From (3.2) the potential of mean force for neighbouring particles reduces to $\phi^{(2)}(r) = \Delta\Omega(r) - 2\mu^{ex}$ if they do not overlap, giving

$$
\begin{aligned}
\phi^{(2)}(r) &\equiv -k_B T \ln g^{(2)}(r) \\
&= pV(r) + \gamma_\infty A(r) + \kappa C(r) + \overline{\kappa} X(r) - 2\mu^{ex} \quad \forall \ \sigma \le r \le 2\sigma.
\end{aligned}
\tag{4.18}
$$

Inserting the morphological quantities at contact from (3.3) gives

$$
\phi^{(2)}(\sigma) = \frac{9\pi\sigma^3}{4}p + 6\pi\sigma^2\,\gamma_\infty + \left(6\pi\sigma - \frac{\pi^2\sigma}{2\sqrt{3}}\right)\kappa + 4\pi\,\overline{\kappa} - 2\mu^{ex}[p].
\tag{4.19}
$$

Equating this with $-k_B T \ln g^{(2)}(\sigma)$ and using the virial theorem (Eq. (4) in the main text) gives the final expression

$$
\frac{9\pi\sigma^3}{4}p + 6\pi\sigma^2\,\gamma_\infty + \left(6\pi\sigma - \frac{\pi^2\sigma}{2\sqrt{3}}\right)\kappa + 4\pi\,\overline{\kappa} = 2\mu^{ex}[p] - \beta^{-1}\ln\frac{3}{2\pi\rho\sigma^3}\left(\frac{\beta p}{\rho} - 1\right).
\tag{4.20}
$$

We will use this last expression instead of the contact theorem (4.15) in order to obtain new coefficients. Together (4.6), (4.7) and (4.20) solve to give coefficients:

$$
\frac{\beta\gamma_\infty^V}{R\rho} = \frac{(18\pi - 7\sqrt{3}\pi^2)p + 6\sqrt{3}\pi\mu^{ex}[p] - 6(12 - \sqrt{3}\pi)\ln(1-\eta) - 72\ln\left(\frac{\pi p - 6\eta}{24\eta^2}\right)}{54(\sqrt{3}\pi - 4)\eta}
\tag{4.21a}
$$

$$
\frac{\beta\kappa^V}{R^2\rho} = \frac{5\pi p - 12\mu^{ex}[p] + 24\ln(1-\eta) + 36\ln\left(\frac{\pi p - 6\eta}{24\eta^2}\right)}{9(\sqrt{3}\pi - 4)\eta}
\tag{4.21b}
$$

$$
\frac{\beta\overline{\kappa}^V}{R^3\rho} = -\frac{(18\pi - 2\sqrt{3}\pi^2)p - (36 - 3\sqrt{3}\pi)\mu^{ex}[p] + 12\sqrt{3}\pi\ln(1-\eta) + 72\ln\left(\frac{\pi p - 6\eta}{24\eta^2}\right)}{27(\sqrt{3}\pi - 4)\eta},
\tag{4.21c}
$$

which upon insertion of the Carnahan-Starling parameters (4.3) and (4.4) gives the coefficients explicitly as

$$
\frac{\beta p^V}{\rho} = \frac{1 + \eta + \eta^2 - \eta^3}{(1-\eta)^3}
\tag{4.22a}
$$

$$
\frac{\beta\gamma_\infty^V}{R\rho} = \frac{18\eta\frac{1+\eta+\eta^2-\eta^3}{(1-\eta)^3} + \sqrt{3}\pi\eta\frac{1-16\eta-4\eta^2+7\eta^3}{(1-\eta)^3} - (12-\sqrt{3}\pi)\ln(1-\eta) - 12\ln\left(\frac{2-\eta}{2(1-\eta)^3}\right)}{9(\sqrt{3}\pi - 4)\eta}
\tag{4.22b}
$$

$$
\frac{\beta\kappa^V}{R^2\rho} = -\frac{2\eta\frac{11-23\eta+\eta^2+5\eta^3}{(1-\eta)^3} - 8\ln(1-\eta) - 12\ln\left(\frac{2-\eta}{2(1-\eta)^3}\right)}{3(\sqrt{3}\pi - 4)\eta}
\tag{4.22c}
$$

$$
\frac{\beta\overline{\kappa}^V}{R^3\rho} = \frac{12\eta\frac{5-12\eta+3\eta^3}{(1-\eta)^3} - \sqrt{3}\pi\eta\frac{4-13\eta-\eta^2+4\eta^3}{(1-\eta)^3} - 4\sqrt{3}\pi\ln(1-\eta) - 24\ln\left(\frac{2-\eta}{2(1-\eta)^3}\right)}{9(\sqrt{3}\pi - 4)\eta}.
\tag{4.22d}
$$

Unlike the WBII coefficients above these are entirely new, and produce significantly more accurate correlation functions at high densities as described in the main text. The pair correlation produced by these coefficients (black line in Fig. 2) is self-consistent with CS at contact by construction, but as an additional bonus we find that the new coefficients provide a theory that outperforms the older WBII approach across the whole range of distances typical of neighbouring particles (inset of Fig. 2). The latter observation enables us to accurately model complex many-particle local structures.

However, it should be noted that the planar surface tension $\gamma_\infty$ is considerably less accurate than $\gamma_\infty^{WBII}$ as compared with molecular dynamics studies in [19]. For this reason, WBII coefficients may give more accurate grand potentials (and thus correlations) for large solutes where the surface becomes approximately planar.

## V. STRUCTURAL FREE ENERGY CALCULATIONS

### A. Structure population integrals and free energies

From the definition of the probability density function in Eq. (1.1), the total number of local structures of a particular type in a volume $V$ is

$$\mathcal{N} = \int_{\mathcal{Q}} \rho^n g^{(n)}(\mathbf{r}^n)\, d\mathbf{r}^n, \tag{5.1}$$

where $\mathcal{Q}$ is the domain *defining* the local structure. To get the free energy expression used in the main text we consider the manifold diffeomorphic to translations, defining $\mathcal{Q} = \mathcal{D} \rtimes \mathbb{V}$ where $\mathbb{V} \subset \mathbb{R}^3$ is the system volume. Exploiting translational invariance of the potential of mean force, we fix one particle at the origin and integrate the center of mass over the system volume giving

$$\mathcal{N} = \rho^n V \int_{\mathcal{D}} g^{(n)}(\mathbf{r}^n)\, d\mathbf{r}^{n-1}. \tag{5.2}$$

We defined a free energy by taking $\mathcal{N} = \sigma^{3(n-1)} \rho^n V e^{-\beta F}$, which together with the above expression gives Eq. (5) in the main text.

More generally, we consider the manifold diffeomorphic to translations *and* rotations. We define $\mathcal{Q} = \mathcal{D}' \rtimes SE(3)$ where SE($d$) is the $d$-dimensional special Euclidean group, leaving the $\mathcal{D}'$ as the space of the structure's internal motion. We separate rigid body from internal motion by applying the following transformation to each particle coordinate

$$\mathbf{r}(\{\mathbf{t}, \boldsymbol{\theta}, \mathbf{x}\}) = \mathbf{t} + \mathbf{R}(\boldsymbol{\theta}) \cdot \mathbf{q}(\mathbf{x}), \tag{5.3}$$

where $\mathbf{t}$ is the translation vector, $\boldsymbol{\theta}$ the Euler angles, $\mathbf{R}$ the rotation matrix, and $\mathbf{x} \in \mathbb{R}^{3n-6}$ represents the internal coordinates. We need to compute the metric of this transformation $G_{ij} = \mathbf{G}_i \mathbf{G}_j^T$ where the (generally curvilinear) basis vectors are $\mathbf{G}_i = \partial_i \mathbf{r}$. To simplify calculation we choose $\mathbf{q}(\mathbf{x})$ to always be in the center-of-mass frame and orthogonal to rotations such that $G_{ij}$ reduces to block-diagonal form. If the rotation matrix is expressed in Euler-angle representation as $\mathbf{R}(\boldsymbol{\theta}) = \mathbf{R}_3(\theta_3)\mathbf{R}_2(\theta_2)\mathbf{R}_1(\theta_1)$ then we have

$$G_{ij}(\{\theta_i\}, \mathbf{x}) = \begin{pmatrix} n\mathbf{E} & 0 & 0 \\ 0 & \mathbf{U}^T(\boldsymbol{\theta})\mathbf{I}(\mathbf{x})\mathbf{U}(\boldsymbol{\theta}) & 0 \\ 0 & 0 & \overline{G_{ij}}(\mathbf{x}) \end{pmatrix}, \tag{5.4}$$

where $\mathbf{E}$ is the identity matrix, $\overline{G_{ij}}$ is the metric for internal motion, and we have defined the matrix $\mathbf{U}$ as

$$\mathbf{U}(\boldsymbol{\theta}) = \begin{pmatrix} 1 & 0 & -\sin\theta_2 \\ 0 & \cos\theta_1 & \cos\theta_2\sin\theta_1 \\ 0 & -\sin\theta_1 & \cos\theta_2\cos\theta_1 \end{pmatrix} \qquad \begin{aligned} &\theta_1 \in [0, 2\pi] \\ &\theta_2 \in \left[-\tfrac{\pi}{2}, \tfrac{\pi}{2}\right] \\ &\theta_3 \in [0, 2\pi]. \end{aligned} \tag{5.5}$$

Note $\det \mathbf{U} = \cos\theta_2$. The volume element in the new coordinates is

$$d\mathbf{r}^n = \frac{\sqrt{\det G_{ij}(\mathbf{x})}}{\nu}\, d^3\mathbf{t}\, d^3\boldsymbol{\theta}\, d^{3n-6}\mathbf{x}, \tag{5.6}$$

where $\nu$ is the symmetry number (discussed below) and

$$\sqrt{\det G_{ij}(\mathbf{x})} = \cos\theta_2 \sqrt{\det \mathbf{I}(\mathbf{x})} \sqrt{n^3 \det \overline{G_{ij}}(\mathbf{x})}. \tag{5.7}$$

The symmetry number emerges as the choice of internal coordinates typically fixes the particle labels breaking permutation symmetry; we have to multiply by the $n!$ possible labellings, which introduces double counting if the structure possesses rotational symmetry so we have to divide by the correcting factor $\nu$. This is explained in detail in Ref. [20]. Thus (5.1) reduces to

$$\frac{\mathcal{N}}{\rho^n V} = \frac{8\pi^2 \sqrt{n^3}}{\nu} \int_{D'} g^{(n)}(\mathbf{x}) \sqrt{\det \overline{G_{ij}}(\mathbf{x}) \det \mathbf{I}(\mathbf{x})}\, d^{3n-6}\mathbf{x}. \tag{5.8}$$

Note that in the limit of linear molecules, i.e. where all particles fall on a line, the above approach fails as there is one less rotation mode requiring a modified description.

In general, the integrand in (5.8) is only exactly solvable for the simplest geometries due to the high dimensionality of $\mathbf{x}$. This is further complicated by the fact that the basis vectors for $\overline{G_{ij}}$ are curvilinear. We use two methods for evaluating these integrands:

1. Monte Carlo simulation: described in the next section.

2. Analytically via perturbation theory: described in sections V C and V D. We use this for a similar integrand along a reaction path where Monte-Carlo cannot be used directly.

## B. Numerical integral with thermodynamic integration

Here we describe the method used in the main text to obtain the free energies of local structure by thermodynamic integration. Aspects of this method were inspired by Ref. [21].

We perform a thermodynamic integration between two potentials $V_1$ and $V_2$ by considering the intermediate potential

$$V(\mathbf{r}^n; V_1, V_2) = \lambda V_1(\mathbf{r}^n) + (1 - \lambda)V_2(\mathbf{r}^n). \tag{5.9}$$

where $\lambda$ takes the allowed values $\{0, 1\}$. Over the course of a Monte-Carlo sweep, in addition to regular steps $\lambda$ is allowed to switch between these values according to the Metropolis-Hastings rule. The free energy difference between the two systems is determined from the ratio of time spent with each potential active by

$$\Delta F \equiv F_2 - F_1 = -\log\left(\frac{t_2}{t_1}\right), \tag{5.10}$$

where $t_i$ is the total number of sweeps where potential $i$ is active (by $\lambda$ value).

Local structure is defined in the main text by its bond topology. We define an $n \times n$ adjacency matrix $\mathcal{A}$ giving the bond topology of $n$ particles by

$$A_{ij} = \begin{cases} 1 & i \text{ and } j \text{ "bonded"} \\ 0 & \text{otherwise} \end{cases} \tag{5.11}$$

The hard core interaction ensures $r_{ij} \equiv |\mathbf{r}_i - \mathbf{r}_j| > \sigma$ for all $(i, j)$ including non-bonded particles, but the "bonded" flag forces the stricter condition that $r_{ij} \in [\sigma, \sigma_{cut}]$ when $A_{ij} = 1$. The latter strict criterion is how we are able to focus on specific local structures.

To do the thermodynamic integration we introduce an intermediate potential which takes into account the boundary conditions of the integral:

$$\phi(\mathbf{r}^n; v_\lambda, \eta) = \sum_{i<j} A_{ij} v_\lambda(r_{ij}) + \phi^{(n)}(\mathbf{r}^n; \eta) \tag{5.12}$$

where $v_\lambda$ is a *confining* potential to restrict the Monte-Carlo to integrate only the structure of interest. We have indicated the dependence of the potential of mean force on volume fraction, noting that in the ideal gas limit the grand potential (solvation) term vanishes leaving only the potential energy contribution

$$\phi^{(n)}(\mathbf{r}^n; \eta = 0) = \sum_{i<j} v_{hs}(r_{ij}), \tag{5.13}$$

where the hard sphere (no-overlap) interaction is

$$v_{hs}(r) = \begin{cases} 0 & r > \sigma \\ \infty & \text{elsewhere.} \end{cases} \tag{5.14}$$

The potential of interest is $\phi(\mathbf{r}^n; v_{hs}, \eta)$, which requires several thermodynamic integration steps to reach.

First, we perform an integration from a reference potential $\phi(\mathbf{r}^n; v_H, \eta = 0)$ to $\phi(\mathbf{r}^n; v_{SW}, \eta = 0)$ where the confining potential $v_\lambda$ takes either the harmonic form

$$v_H(r) = \epsilon \left(\frac{\sigma_{cut} - r}{\sigma}\right)^2 \tag{5.15}$$

with a value of $\epsilon$ discussed below, or the (infinite) square well form

$$v_{SW}(r) = \begin{cases} 0 & \sigma < r < \sigma_{cut} \\ \infty & \text{elsewhere} \end{cases} \tag{5.16}$$

to properly impose the boundary conditions $r_{ij} \in [\sigma, \sigma_{cut}]$ when $A_{ij} = 1$. To optimise the simulations the value of $\epsilon$ should be chosen to keep the free energy difference between the two systems of order $\mathcal{O}(1\,k_BT)$; we found $\epsilon = 75\sigma^2$ to be a reasonable choice for cutoff $\sigma_{cut} = 1.2\sigma$ at the system sizes considered ($n \leq 12$). The free energy of the reference harmonic system is found by multivariate Gaussian integration, neglecting the effect of the hard sphere interactions for $\epsilon \gg 1$. In subsequent steps we integrate between $\phi(\mathbf{r}^n; v_{SW}, \eta_i)$ and $\phi(\mathbf{r}^n; v_{SW}, \eta_{i+1})$ until the desired final volume fraction is reached.

### C. Analytical integral in bond distance space

To evaluate the integrand in Eq. (5.8) analytically we need to choose a representation for $\mathbf{x}$ which is diffeomorphic to $\mathbf{r}^n$. For *minimally constrained geometries*, i.e. structures with exactly $3n - 6$ contacts, a convenient representation exists: bond distance space. Following Ref. [22] and its accompanying Supplementary Information we choose each element of $\mathbf{x}$ to represent the distances between particles in contact, where contact occurs at $\mathbf{x} = (\sigma, \ldots, \sigma)$. Thus increasing elements of $\mathbf{x}$ corresponds to thermal fluctuations away from contact. This representation naturally expresses the limits of integration given in the main text as $\sigma \leq x_i \leq \sigma_{cut}$.

To evaluate the integral we need expressions for the internal metric and moment of inertia terms. The internal metric is defined in terms of the Jacobian matrix $\mathbf{J} \in \mathbb{R}^{3n \times 3n - 6}$

$$\overline{G_{ij}} = \mathbf{J}^T \mathbf{J} \tag{5.17}$$

where the matrix entries are given by

$$J_{ij} = \frac{\partial q_i}{\partial y_j}. \tag{5.18}$$

In practice it is easier to calculate its inverse numerically (via finite differences) using

$$J_{ij}^{-1} = \frac{\partial y_j}{\partial q_i} = \frac{\partial y_j}{\partial x_i}\bigg|_{\boldsymbol{\theta} = \mathbf{t} = 0} \tag{5.19}$$

which has linearly independent rows for a minimally constrained geometry so we recover $\mathbf{J}$ from $\mathbf{K} = \mathbf{J}^{-1}$ using the matrix inversion formula $\mathbf{J} = (\mathbf{K}^T\mathbf{K})^{-1}\mathbf{K}^T$. The above expressions all depend implicitly on the point $\mathbf{x}$ as the coordinate scheme is curvilinear, so to keep the integral tractable we approximate this to leading order as

$$\sqrt{\overline{G_{ij}(\mathbf{x})}} \simeq \sqrt{\overline{G_{ij}(\mathbf{x}_0)}} \tag{5.20}$$

where $\mathbf{x}_0 = (\sigma, \ldots, \sigma)$ is contact. Thus the integral becomes

$$\frac{\mathcal{N}}{\rho^n V} = \frac{8\pi^2 G_0}{\nu} \int_{D'} g^{(n)}(\mathbf{x})\, \sqrt{\det \mathbf{I}(\mathbf{x})}\, d^{3n-6}\mathbf{x} \tag{5.21}$$

where $G_0 = \sqrt{n^3 \det \overline{G_{ij}(\mathbf{x}_0)}}$.

Finally, we write the distribution function in terms of the potential of mean force and expand this and the moment of inertia to first order, as in

$$\phi^{(n)}(\mathbf{x}) = \phi^{(n)}(\mathbf{x}_0) + (\mathbf{x} - \mathbf{x}_0) \cdot \boldsymbol{\nabla}\phi^{(n)}(\mathbf{x})\big|_{\mathbf{x} = \mathbf{x}_0} + \mathcal{O}(\mathbf{x}^2), \tag{5.22}$$

$$\sqrt{\det \mathbf{I}(\mathbf{x})} = \sqrt{\det \mathbf{I}(\mathbf{x}_0)} + (\mathbf{x} - \mathbf{x}_0) \cdot \boldsymbol{\nabla}\sqrt{\det \mathbf{I}(\mathbf{x})}\big|_{\mathbf{x} = \mathbf{x}_0} + \mathcal{O}(\mathbf{x}^2). \tag{5.23}$$

Using the analytical gradient expressions given in Section II makes this calculation very efficient. The integral (5.21) separates into $3n - 6$ independent one-dimensional integrals of the form

$$\int_{\sigma}^{\sigma_{cut}} (a_i + b_i x_i) e^{-c_i x_i} dx_i = \left[ -\left( \frac{(a + b_i x_i)}{c_i} + \frac{b_i}{c_i^2} \right) e^{-c_i x_i} \right]_{\sigma}^{\sigma_{cut}},$$

where $a_i$, $b_i$ and $c_i$ are constants.

Loosely speaking, this is the hard-particle analogue of the harmonic approximation with the difference here being that the first derivative does not vanish at the minimum. For $n = 6$ this expansion works rather well, as all structures have exactly $3n - 6$ bonds and this perturbation theory captures the free energy well when compared with the "exact" result from thermodynamic integration.

### D. Integrating along reaction paths

For $n < 6$ only a single rigid packing of hard spheres exists. For $n = 6$ we see two distinct packings making it the first interesting landscape. The two structures are connected by a single unstable reaction path, making the entire landscape simple enough to explore. However, to evaluate free energies along this path we cannot use the thermodynamic integration method from section V B as the Monte Carlo procedure will cause sampling along the unstable direction which must be excluded. As such we require an analytic approach, for which we will use the geometry of bond distance space integration method described in the previous section V C.

From (5.21) we have the free energy of a local structure as

$$\beta F = -\ln \frac{\mathcal{N}}{\rho^n V \sigma^{3(n-1)}} = -\ln \left( \frac{8\pi^2 G_0}{\nu \sigma^{3(n-1)}} \int_{D'} g^{(n)}(\mathbf{x}) \sqrt{\det \mathbf{I}(\mathbf{x})} \, d^{3n-6}\mathbf{x} \right) \tag{5.24}$$

where the boundary conditions in bond-distance space correspond to the hypercube $\sigma \le x_i \le \sigma_{cut}$. Now we generalise this result to evaluate free energies of intermediate states, by parameterising the intermediate state in terms of one broken bond $x_b$. We therefore obtain

$$\beta F = -\ln \frac{\mathcal{N}}{\rho^n V \sigma^{3(n-1)}} = -\ln \left( \frac{8\pi^2 G_0}{\nu_{\mathrm{RP}} \sigma^{3(n-1)}} \int_{D'_{\mathrm{RP}}} g^{(n)}(\mathbf{z}; x_b) \sqrt{\det \mathbf{I}(\mathbf{z})} \, d^{3n-7}\mathbf{z} \right) \tag{5.25}$$

where here $\mathbf{z} \in \mathbb{R}^{(3n-7)}$ does not include bond $z_n$, and $\nu_{RP}$ is the new symmetry number along this reaction path, which is in general different from the symmetry number at the terminating stationary points [23]. We proceed to evaluate the integral (5.25) by the same method as outlined in the previous section, i.e. by Taylor expansion up to linear terms in the potential and inertia so the integral reduces to a sum of one-dimensional integrals.

## VI. BULK MOLECULAR DYNAMICS

Molecular dynamics simulations of bulk systems were used in the main text to verify the effectiveness of the theory in predicting populations of local structure. We used event-driven molecular dynamics of $N = 1372$ hard spheres using the DynamO software package [24] with two size distributions:

1. A single component system for densities below freezing $\eta \le 0.49$.

2. A 5-component equimolar distribution with $\sim 8\%$ polydispersity, to avoid freezing at high densities. Data is unavailable at very high densities $\eta \gtrsim 0.58$ where the system is so dynamically arrested that equilibration is impossible within simulation timescales.

Equilibrium was achieved by running the simulations for $\sim 100$ ($\alpha-$) relaxation times. The resulting equilibrium configurations were analysed for structure using the topological cluster classification [25]. Neighbour detection used a simple pairwise cutoff to be consistent with the structural definitions used by the theory described in the main text.

---

%

\end{document}